\documentclass[aps,pra,twocolumn,superscriptaddress]{revtex4-1}

\usepackage{graphicx}
\usepackage{hyperref}
\usepackage{amsmath}
\usepackage{amssymb}
\usepackage{epstopdf}
\usepackage{multirow}
\usepackage{threeparttable}
\usepackage[usenames]{color}

\begin{document}

\title{Mid-infrared nonlinear pinhole imaging}
\author{Yanan Li}
\affiliation{State Key Laboratory of Precision Spectroscopy, and Hainan Institute, East China Normal University, Shanghai 200062, China}

\author{Kun Huang}
\email{khuang@lps.ecnu.edu.cn}
\affiliation{State Key Laboratory of Precision Spectroscopy, and Hainan Institute, East China Normal University, Shanghai 200062, China}
\affiliation{Chongqing Key Laboratory of Precision Optics, Chongqing Institute of East China Normal University, Chongqing 401121, China}
\affiliation{Collaborative Innovation Center of Extreme Optics, Shanxi University, Taiyuan, Shanxi 030006, China}

\author{Jianan Fang}
\affiliation{State Key Laboratory of Precision Spectroscopy, and Hainan Institute, East China Normal University, Shanghai 200062, China}
\affiliation{Chongqing Key Laboratory of Precision Optics, Chongqing Institute of East China Normal University, Chongqing 401121, China}

\author{Zhuohang Wei}
\affiliation{State Key Laboratory of Precision Spectroscopy, and Hainan Institute, East China Normal University, Shanghai 200062, China}

\author{Heping Zeng}
\email{hpzeng@phy.ecnu.edu.cn}
\affiliation{State Key Laboratory of Precision Spectroscopy, and Hainan Institute, East China Normal University, Shanghai 200062, China}
\affiliation{Chongqing Key Laboratory of Precision Optics, Chongqing Institute of East China Normal University, Chongqing 401121, China}
\affiliation{Shanghai Research Center for Quantum Sciences, Shanghai 201315, China}
\affiliation{Chongqing Institute for Brain and Intelligence, Guangyang Bay Laboratory, Chongqing, 400064, China}

\begin{abstract}
Pinhole imaging is the most primitive and simplest lensless imaging paradigm, capable of transcending the physical limitations of conventional lens optics. This modality is particularly attractive for accessing a virtually infinite depth of focus or operating at extreme wavelengths. Here, we devise and implement a mid-infrared (MIR) pinhole imaging system at 3.07 $\mu$m based on nonlinear spatial filtering. Instead of using a physical aperture, the involved pinhole is optically formed by a near-infrared pump at 1.03 $\mu$m within a nonlinear crystal, which allows flexible and precise control over the effective aperture size to optimize imaging performance. Meanwhile, the MIR rays passing through the nonlinear pinhole are spectrally upconverted to facilitate sensitive imaging via a silicon camera. Consequently, the implemented upconversion pinhole imaging enables a large depth of field over 35 cm, beyond the reach of typical lens-based upconversion imagers. Furthermore, depth-resolving imaging across a large depth range is demonstrated in both the reflection and transmission modes based on time-of-flight and trigonometric techniques, respectively. The achieved capabilities---featuring large operation depth, wide field of view, and flexible adaptability to various illumination conditions---highlight the potential of the presented MIR imaging architecture for expansive scene detection and motion-aware applications in industrial inspection and night vision.
\end{abstract}

\maketitle

\section{Introduction}
Lensless imaging is a powerful approach in modern optics and computational imaging, which has facilitated various applications including biological microscopy, photography, endoscopy, machine vision \cite{Ozcan2016ARBE, Li2024FR}. In the lensless scheme, optical wavefront modulation is typically used along with computational algorithms to achieve complex scene sensing and reconstruction \cite{Boominathan2022Optica}. Lensless designs usually encompass holographic lensless imaging, mask-modulated imaging, and illumination-modulated imaging \cite{Boominathan2022Optica, Potter2024LPR, Antipa2018Optica}. With the rapid development of novel modulation devices and artificial intelligence algorithms, novel computational lensless paradigms have emerged, such as programmable mask imaging \cite{Hua2020IEEE TPAMI} and deep learning-based imaging \cite{Sinha2017Optica, Wu2024LSA}. These advancements not only significantly improve imaging performance, but also enable simultaneous extraction of multimodal data including phase, polarization, and spectral information \cite{Monakhova2020Optica, Baek2022APLP}.

Particularly, pinhole imaging represents the most primitive and fundamental lensless imaging modality \cite{Liang2020RPP, Cieslak2016RM}. Early recorded descriptions can be traced back to pioneering works by Mozi (c. 370 B.C.) and Alhazen (965-1039 A.D.) \cite{Boominathan2016IEEE SPM}. The principle relies on the rectilinear propagation of light through a tiny aperture, forming an inverted real image without the use of any optical lens. Despite its conceptual simplicity, pinhole imaging offers distinct advantages: it is inherently free from linear distortion, provides an effectively infinite depth of field, and supports a wide angular field of view \cite{Gong2009APL, Wang2015OC}. Continued exploration of pinhole imaging has led to the development of the pinhole camera, or camera obscura \cite{Franke1979AO, Lindberg1970AHES}. Originally applied in early photography, the pinhole imaging also laid the technical foundation for innovations in perspective drawing during the Renaissance \cite{Young1989PT}. Subsequently, it has stimulated widespread applications in astronomical observation \cite{Straker1981AHES}, X-ray imaging \cite{Biri2011IEEE TPS}, flight simulation \cite{Gallas1965JSMPTE}, integrated circuit manufacturing \cite{Newman1966AO}, and science education \cite{Young1971AO}.

To date, the pinhole imaging predominantly operates at visible wavelengths, further extension to mid-infrared (MIR) region has attracted particular interest due to chemical specificity and thermal sensitivity \cite{Vodopyanov2020Book}. However, the inherently low light throughput of pinholes necessitates long exposure time and highly sensitive detection \cite{Young1989PT}, posing a significant challenge to realize MIR pinhole imaging. Indeed, MIR cameras based on the narrow-bandgap semiconductors like HgCdTe and InSb \cite{Rogalski2005RPP}, have long been plagued by high dark noise, low pixel count, and thermal susceptibility \cite{Wang2019Small}. The use of cryogenic operation can enhance detection sensitivity, such as superconducting nanowires and transition-edge sensors \cite{Taylor2023Optica}, albeit with additional complexity of the cooling system. Notably, emerging low-dimensional materials, like colloidal quantum dots \cite{Keuleyan2011NP}, black phosphorus \cite{Bullock2018NP}, graphene \cite{Guo2018NM}, and tellurium nanosheet \cite{Peng2021SA}, show great promise for sensing infrared photons at room temperature, yet large-area deposition for high-pixel-density arrays are still in the infancy \cite{Liu2021LSA}. Given the advantages of high-performance silicon-based cameras, upconversion strategies that translate MIR signals into shorter wavelengths have been widely recognized as an indirect yet effective approach for MIR imaging \cite{Barh2019AOP, Dam2012NP, Huang2022NC, Zeng2023LPR}. Various platforms have been explored to realize this process, including nonlinear crystals \cite{Wang2023NC, Ge2023PRAppl, Mrejen2020LPR}, metasurfaces \cite{Morales2021AP}, fluorophores \cite{Zheng2013NP}, and two-photon absorption \cite{Knez2020LSA}. By integrating the techniques of upconversion detection and pinhole imaging, it becomes possible to synergistically harness the high detection sensitivity of the upconverted imager while mitigating the limitations of lens-based imaging systems, such as optical aberrations and restricted depth of field.

Here, we propose and demonstrate a lensless MIR nonlinear pinhole imaging scheme at 3.07 $\mu$m. The involved pinhole is optically formed by the 1.03-$\mu$m pump beam rather than a physical aperture. This nonlinear optical pinhole enables flexible and precise spatiotemporal control, serving simultaneously as a spatial filter for image formation and as a pump source for frequency upconversion. Compared to conventional lens-based MIR upconverters, the proposed scheme provides broader imaging coverage, supporting a depth range exceeding 35 cm and a field of view over 6 cm. Notably, the synchronized pump pulse in the nonlinear process facilitates an ultrashort optical gate in time-of-flight imaging, where the noise is substantially suppressed within a narrow temporal window to support a detection sensitivity about 1.5 photons/pulse. Furthermore, we introduce a transmissive depth-resolving imaging modality, capable of rapid scene reconstruction even under passive illumination, thus opening new opportunities for MIR sensing and imaging. Note that the operation MIR wavelength has been chosen to demonstrate the system's effectiveness, and could be extended to cover a wide spectral window of 3-5 $\mu$m thanks to the broadband nonlinear upconverter.

\begin{figure*}[t!]
\includegraphics[width=0.85\textwidth]{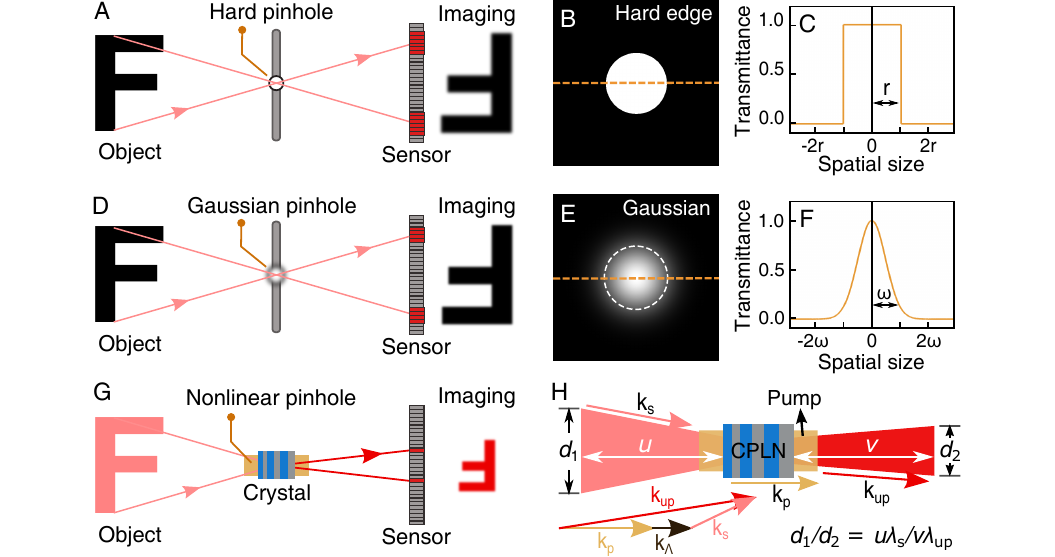}
\caption{Conceptual illustration for MIR nonlinear-pinhole imaging. (A) Schematic diagram of traditional pinhole imaging. (B) Aperture shape and intensity distribution in traditional pinhole imaging systems. (C) Corresponding cross section along the dashed line in (B). (D) Schematic diagram of Gaussian pinhole imaging, where the softened aperture edge reduces the diffraction effect, thus improving the imaging resolution. (E) Intensity distribution of the Gaussian aperture, with the white dashed circle corresponding to the $1/{e^2}$ radius. (F) Corresponding cross section along the dashed line in (E). (G) Schematic diagram of MIR nonlinear optical pinhole imaging, where the pump light within the nonlinear crystal is equivalent to the optical aperture. (H) Phase-matching condition for the MIR upconversion imaging. The deflection angle of the upconverted beam depends on the signal wavelength, thus affecting the image scaling factor. $u$ and $v$ represent the distances of the object and the camera from the optical aperture, respectively. $d_1$ and $d_2$ represent the sizes of the object and the camera probe image. $\lambda_\text{s}$ and $\lambda_\text{up}$ represent the wavelengths in vacuum for the signal and upconverted light, respectively.}
\label{fig1}
\end{figure*}

\section{Basic principle}
The MIR nonlinear pinhole imaging is essentially a lensless technique that generates upconverted images based on the principle of pinhole imaging. Traditional MIR imaging relies on lens refraction and requires precise object-to-lens distances constrained by the focal length, thereby limiting the depth of field (DoF). In contrast, pinhole imaging enables image formation across an effectively infinite depth range. As illustrated in Fig. \ref{fig1}(A), pinhole imaging refers to the modality where light rays emitted from an object pass through a small aperture, forming an inverted real image on the sensor plane. The relationship between the image and object sizes is given by $M = -{d_2}/{d_1}$, where $M$ is the magnification, and $d_{1,2}$ are the heights of the object and the image. The negative sign indicates that the image is inverted. A key characteristic of pinhole imaging is that the pinhole acts as a directional filter for light rays. Without the pinhole, each point on the screen would be illuminated by light from a broad angular range, resulting in a uniformly lit spot. In contrast, with a pinhole in place, only light from specific directions reaches each point on the screen, forming an inverted image of the scene \cite{Tomasi2015NCS}. A classic pinhole refers to a physical aperture with well-defined edges, where light intensity drops sharply near the boundary as shown in Figs. \ref{fig1}(B) and (C). The radius $r$ is defined as the distance from the center to the edge of the hole.

Beyond simply adjusting aperture size, imaging quality can also be enhanced by shaping the transmittance profile of the pinhole. Figures \ref{fig1}(D-F) illustrate a Gaussian pinhole, where the transmitted intensity decreases gradually from the center outward. The radius $\omega$ of the Gaussian pinhole is defined as the distance from the center to the point where the intensity falls to $1/{e^2}$ of its peak value. The Gaussian profile provides a smoother intensity transition at the aperture edge, thereby reducing diffraction effects and improving imaging resolution in the case of small apertures \cite{Liu2020AO}. More details are given in Supplement 1, Note 2.

In our proposed scheme, MIR upconversion imaging is realized by constructing an optical pinhole with a Gaussian intensity profile during the nonlinear frequency conversion process as depicted in Fig. \ref{fig1}(G). This nonlinear optical pinhole, formed by the pump beam within the nonlinear crystal, provides both flexibility and precision in spatial control. It simultaneously functions as the pump source for upconversion and as the optical aperture for image formation. Notably, the nonlinear crystal directly receives MIR signal light from the object, in contrast to conventional MIR upconversion systems that rely on 2f or 4f lens configurations. This lensless operation circumvents the DoF limitations imposed by lenses, while efficiently converting MIR signals into near-infrared or visible wavelengths. Therefore, it could enable sensitive and wide-depth MIR imaging.

Moreover, the nonlinear imaging process induces a deflection in the emission angle due to the wavelength shift of the upconverted signal, resulting in a variation in the image-to-object magnification ratio. This behavior arises from the sum-frequency generation (SFG) process, which satisfies the phase-matching condition, as illustrated in Fig. \ref{fig1}(H). Efficient frequency conversion requires momentum conservation in the transverse direction, $\textit{i.e.}$, minimizing the phase mismatch $\Delta k_{\perp} = k_\text{up} \sin \theta_\text{up} - k_\text{s} \sin \theta_\text{s}$, where $k_{\text{s},\text{up}}$ are wave vectors of the signal and upconverted light inside the crystal, and $\theta_{\text{s},\text{up}}$ are their respective propagation angles relative to the optical axis. By applying Snell's law at the air-crystal interface, the angular magnification between the external angles $\theta'_\text{s}$ and  $\theta'_\text{up}$is given by ${\sin \theta'_\text{up}}/{\sin \theta'_\text{s}} = {\lambda_\text{up}}/{\lambda_\text{s}}$, where $\lambda_\text{s}$ and $\lambda_\text{up}$ are signal and SFG wavelengths in vacuum, respectively \cite{Huang2022NC}. Under the small-angle approximation, the image magnification can be expressed as:
\begin{equation}
M = - \frac{d_2}{d_1} \approx - \frac{v\sin \theta'_\text{up}}{u\sin \theta'_\text{s}} = - \frac{v\lambda_\text{up}}{u\lambda_\text{s}} \ ,
\label{eq1}
\end{equation}
where $u$, $v$ represent the object and image distances from the optical aperture. The energy conservation condition for SFG is given by ${1}/{\lambda_\text{up}} = {1}/{\lambda_\text{s}} + {1}/{\lambda_\text{p}}$, where $\lambda_\text{p}$ is the vacuum wavelength of the pump light.

\begin{figure*}[t!]	
\includegraphics[width=0.85\textwidth]{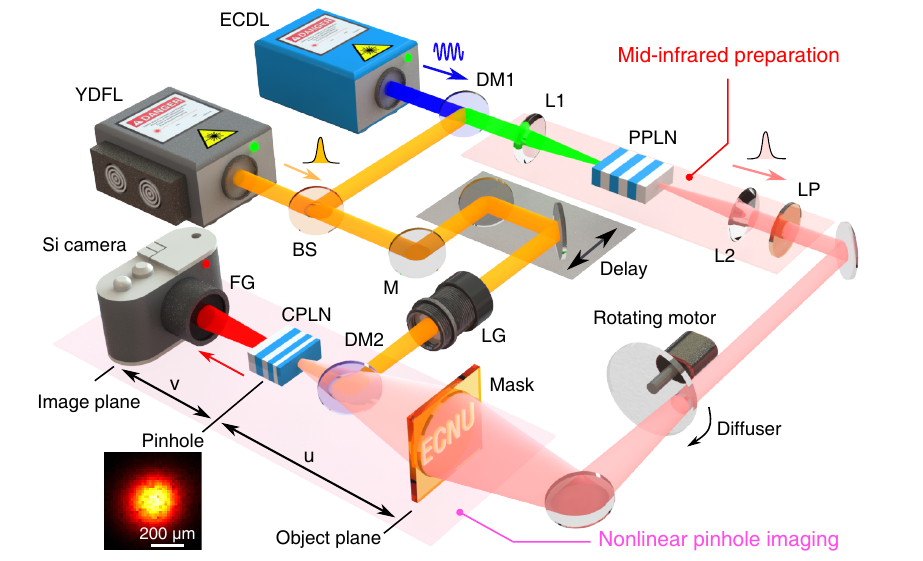}
\caption{Experimental setup of the MIR nonlinear-pinhole upconversion imaging system. The laser sources consist of a femtosecond ytterbium-doped fiber laser (YDFL) and an extended cavity diode laser (ECDL). The YDFL produces mode-locked pulses at 1030 nm, while the ECDL operates in the continuous-wave mode at 1550 nm. The two beams are spatially combined by a dichroic mirror (DM1) and focused into a periodically poled lithium niobate (PPLN) crystal to generate MIR pulses via difference-frequency generation (DFG). The MIR beam passes through a spinning ground-glass diffuser to reduce spatial coherence before illuminating the object. The transmitted (or reflected) MIR signal is then spatially combined with the pulsed YDFL beam at DM2, and temporal overlap is adjusted via a delay line. The combined beams enter a chirped-poling lithium niobate (CPLN) crystal, where the YDFL pulse acts simultaneously as a pump for nonlinear upconversion and as an optical pinhole for the image mapping. After a spectral filtering group (FG), the upconverted image is recorded by a silicon-based camera. Inset at the left-bottom corner gives the pump beam in the crystal, which serves as the optical pinhole. Notably, the illumination could be configured in reflective and transmissive modalities. Both allow the depth-resolving imaging. L: lens; LG: lens group; LP: long-pass filter; BS: beam splitter;  M: silver mirror.}
\label{fig2}
\end{figure*}

\section{Experimental setup}
Figure \ref{fig2} illustrates the experimental setup for MIR nonlinear-pinhole upconversion imaging, which includes both synchronized MIR pulse generation and nonlinear pinhole imaging. The light sources consist of an ytterbium-doped fiber laser (YDFL) and an extended cavity diode laser (ECDL). The YDFL is a mode-locked ultrafast laser operating at 1030 nm, delivering 174 fs pulses at a repetition rate of 20.25 MHz. These ultrashort pulses not only enhance the efficiency of subsequent nonlinear interactions due to their high peak power but also suppress background noise through temporal gating. A portion of the YDFL output is spatially combined with the continuous-wave beam at 1550 nm from the ECDL. The combined beams are then focused into a periodically poled lithium niobate (PPLN) crystal using an achromatic lens to facilitate difference-frequency generation (DFG), producing MIR pulses centered at 3070 nm. The generated MIR beam is collimated by a calcium fluoride lens (LBTEK, MCX70612), and then filtered via a 2.4 $\mu$m long-pass filter to remove unwanted spectral components arising from other nonlinear processes. The resulting MIR pulses are passively synchronized with the pump pulses. These synchronized pulses at two distinct wavelengths are crucial for implementing coincidence-pumped upconversion imaging, which enables efficient and temporally selective signal detection.

Subsequently, the prepared MIR signal, along with the remaining portion of the YDFL output, is directed into the nonlinear pinhole imaging section. The MIR beam passes through a spinning ground-glass diffuser to reduce its spatial coherence. The transmitted image of the object is focused into the center of a chirped-poling lithium niobate (CPLN) crystal. Note that the system also supports the reflective illumination modality, wherein diffused MIR photons reflected from the object are collected and directed into the crystal (see Fig. \ref{fig4}(A)). In parallel, the other portion of the YDFL output serves as the pump source. After passing through a delay line, the pump pulse is spatially combined with the MIR signal to perform the SFG in the CPLN crystal. The pump light forms an optical aperture to facilitate the MIR nonlinear pinhole imaging. The pump beam size can be flexibly adjusted using a lens group to optimize imaging performance. Unlike a conventional physical pinhole, the nonlinear optical pinhole is defined by a three-dimensional focal volume of the pump beam within the nonlinear crystal. This volumetric interaction region determines the effective spatial filtering in both lateral and axial dimensions, influencing image resolution and depth selectivity. The CPLN crystal used has dimensions of 3$\times$2$\times$5 mm$^3$  (width$\times$thickness$\times$length), with a poling period that linearly varies from 16 to 24 $\mu$m along the axial direction. While the spatial confinement of the pump beam defines the effective pinhole aperture, the use of a chirped-period structure is crucial for enabling wide-angle phase matching. It ensures that MIR rays from various incident angles can be efficiently upconverted across an extended field of view. Moreover, the 5-mm crystal length provides sufficient interaction distance to maintain high SFG efficiency within the spatially localized pump-defined aperture. The upconverted image at 771 nm is spectrally filtered to eliminate parametric fluorescences and ambient noises. The filtered signal is finally detected by a silicon-based electron-multiplying CCD (EMCCD, Andor iXon Ultra 888). The conversion efficiency from MIR to visible is measured to be 0.16\%, which is comparable to previous report for wide-field upconversion imaging based on CPLN crystal \cite{Huang2022NC}. The efficiency can be further improved by increasing the pump average power to boost peak intensity. The implemented MIR nonlinear pinhole imaging system allows flexible control over the object-to-crystal distance, the crystal-to-image distance, and the aperture size, thus enabling panoramic DoF imaging. The combination of low-noise nonlinear conversion and high-sensitivity photon detection is critical for realizing low-light-level MIR upconversion pinhole imaging.

\begin{figure*}[t!]
	\includegraphics[width=0.8\textwidth]{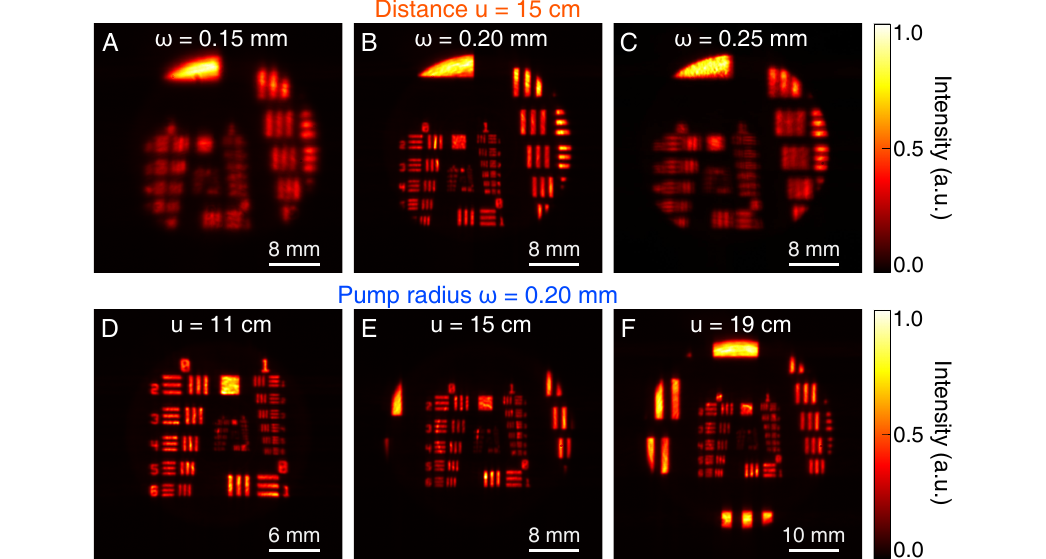}
	\caption{Performance of the MIR nonlinear pinhole imaging. (A-C) Upconverted images acquired at an object distance of 15 cm in the case of various pump beam radii $\omega$ of 0.15 mm, 0.20 mm, and 0.25 mm, respectively. (D-F) Recorded upconverted images with a fixed pump radius of 0.2 mm at different object distances of 11 cm, 15 cm, and 19 cm, respectively. As expected, an increase in the object distance leads to an expanded imaging field of view.}
	\label{fig3}
\end{figure*}

\section{Results and discussion}
\subsection{Characterization of MIR pinhole imaging}
Now we turn  to evaluate the performance of the implemented MIR nonlinear pinhole imaging system. Compared to lens-based approaches, pinhole-based MIR upconversion imaging offers several distinct advantages, including complete freedom from linear distortion, an effectively infinite depth of field, and a wide angular field of view, particularly in scenarios where spatial resolution is not the primary concern \cite{Young1989PT}. However, an infinite DoF does not imply the absence of optical blurring. Rather, it means that blurring is decoupled from object distance and instead governed by other factors---most notably, the aperture size. A larger aperture admits more light, but also increases the likelihood of overlapping light rays, resulting in image blur. Conversely, a smaller aperture restricts light to reduce the overlap,  but the enhanced diffraction would also lead to blurring. Therefore, optimizing the nonlinear optical pinhole with a proper aperture size is essential for achieving high-quality imaging performance. Related behaviors have been theoretical modeled in Supplement 1, Note 1.

\begin{figure*}[t!]
\includegraphics[width=0.85\textwidth]{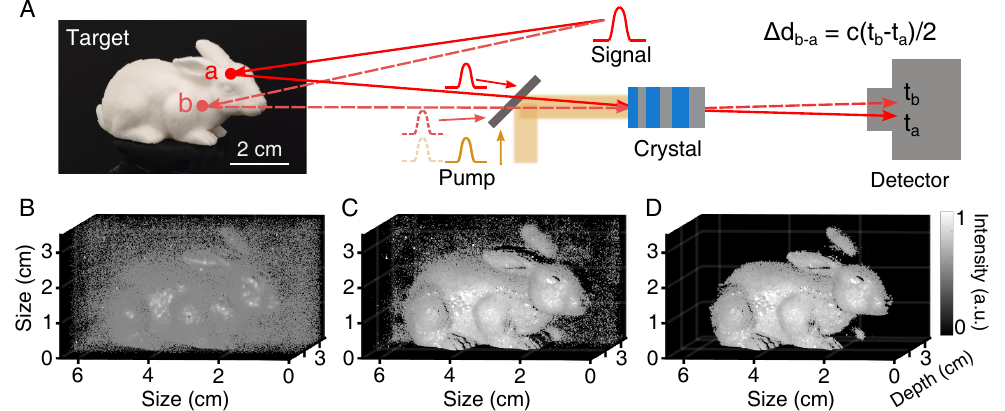}
\caption{Three-dimensional MIR pinhole imaging. (A) The time-of-flight method is used in the reflective illumination mode for performing the three-dimensional imaging. The ground truth target is a matte ceramic rabbit. The relative depth $\Delta d_{b-a}=c(t_b - t_a)/2$ between points $a$ and $b$ is derived from the time difference $t_b - t_a$ of the returned signals, where $c$ is the speed of light. (B) Initial 3D reconstruction obtained by selecting the intensity peak positions along the temporal axis across 170 captured frames. (C) Refined point cloud after applying a spatio-temporal correlation-enhancement algorithm for the noise suppression. (D) Final 3D reconstruction after an edge-based denoising operation, highlighting structural contours of the target.}	
\label{fig4}
\end{figure*}

We evaluate the DoF performance of the system using a standard USAF 1951 resolution target. The imaging distance is fixed at $v=20$ cm, and a diaphragm is placed in front of the camera to maintain a consistent receiving angle for direct comparison. The size of the nonlinear optical aperture is adjusted by tuning the pump beam diameter within the nonlinear crystal via a lens group. Figures \ref{fig3}(A-C) show the evolution of the upconverted images as the aperture radius $\omega$ increases from 0.15 mm to 0.25 mm at an object distance of 15 cm. The image in Fig. \ref{fig3}(B) exhibits high sharpness and minimal distortion. The optimal pinhole width reflects a balance between diffraction and geometric blur. Smaller apertures suffer from increased diffraction, while larger apertures degrade resolution through overlapping rays. The chosen value corresponds to the condition where the total blur is minimized, yielding the best spatial resolution under the system geometry. It is important to note that the optimized value of $\omega$ depends on both object and image positions, and is shown here to illustrate relative trends.  

After determining the optimal aperture, we vary the object distance $u$ to investigate the imaging performance. Figures \ref{fig3}(D-F) correspond to object distances of 11 cm, 15 cm, and 19 cm with a radius $\omega$ of 0.2 mm, respectively. All resulting upconverted images maintain clear and sharp detail, which is beyond the reach of conventional lens-based systems. To quantitatively assess the imaging performance, we evaluate the modulation transfer function (MTF). For patterns in Group 0, Element 1 (500-$\mu$m bars), the MTF peaks at 0.81 when the object is at 15 cm, and decreases to 0.74 at 11 cm and 0.59 at 19 cm. For a higher spatial frequency pattern (Group 0, Element 6; 281-$\mu$m bars), the MTF is 0.45 at 15 cm, dropping to 0.37 and 0.21 at 11 cm and 19 cm, respectively. These results confirm a depth-dependent resolution profile with optimal spatial fidelity at a specific object plane, while maintaining stable contrast across a wide axial range, which is consistent with the large depth of field inherent to pinhole-based imaging.

Performance comparison for various imaging schemes is given in Supplement 1, Note 3. It can be seen that the field of view of the recorded images expands nearly linearly as the object distance $u$ increases. This behavior aligns with the previously described geometric scaling law in Eq. \eqref{eq1}. In addition, the system allows fine-tuning of the aperture to achieve optimal image quality at various object distances, thereby highlighting its effectively infinite depth of field. This capability for large-DoF imaging at a long distance holds significant potential for applications related to infrared monitoring and surveillance. A demonstration of this feature, featuring the movement of an ``ECNU" mask, is provided in Visualization 1.

Although the imaging system is lensless, the effective $F$-number can be defined as $F/\# = v/(2\omega)$. The effective numerical aperture (NA) can be estimated as $\text{NA} \approx 1/(2 \times F/\#)$. In our implementation, with a millimeter-scale pinhole and an image distance of 20 cm, the resulting values are $F/\# \approx 200$ and $\text{NA} \approx 0.0025$. While this is much smaller than the NA of typical MIR lens-based systems, it reflects the system's design tradeoff: reduced light throughput in exchange for an extremely large depth of field and strong spatial selectivity. These characteristics are intrinsic to pinhole-based imaging and are well compensated by the system's high sensitivity and low-noise upconversion detection scheme.

\subsection{Three-dimensional MIR pinhole imaging}
Next, we investigate the three-dimensional (3D) imaging capability of the MIR pinhole imager. As shown in Fig. \ref{fig4}(A), a reflective illumination is configured to implement the time-of-flight 3D imaging for a ceramic rabbit. The MIR pulse illuminates the object, and back-scattered photons from surfaces at different depths coincide temporally with synchronized pump pulses \cite{Gariepy2015NC}. The pump pulse serves as an ultrafast optical gate to time-stamp the signals, producing a data cube containing both spatial and temporal information \cite{Fang2023LSA}. This allows extraction of 2D spatial information via point-to-point mapping between object and image planes, while depth information, specifically the relative depth $\Delta d_{b-a}$, is retrieved from the time delay $t_b - t_a$ between pulses originating from positions $a$ and $b$. As the interaction is governed by the arrival time of the pulses, the measurable depth range depends on the delay scanning window. Ultimately, the maximum unambiguous measurable range is determined by the pulse interval. The resolution of depth measurement is governed by the pulse durations of the involved synchronized lasers. Thanks to the ultrashort pulse used in our experiment, temporal precision at the femtosecond level is achievable, corresponding to $\mu$m-scale depth precision. Further discussion of the achievable depth resolution is provided in Supplement 1, Note 4.

\begin{figure*}[t!]
	\includegraphics[width=0.85\textwidth]{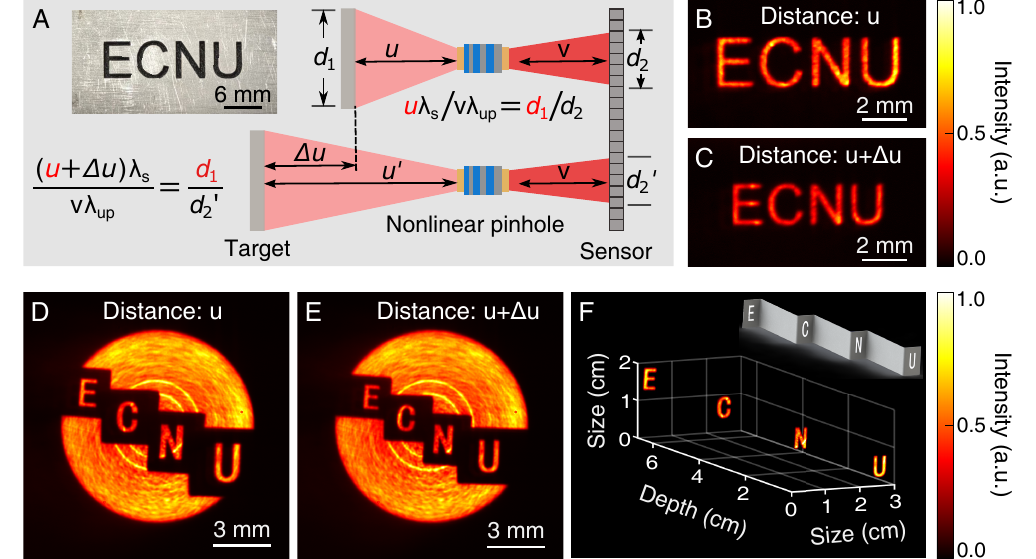}
	\caption{Depth-resolved MIR pinhole photography. (A) Principle of photographic depth-resolved reconstruction. By capturing two images of an object at different object distances separated by $\Delta u$, the spatial dimensions and axial (depth) information can be retrieved from the object-image triangular geometry. (B, C) Captured upconverted images of a planar object at an initial distance $u$ and after shifting it by $\Delta u$, respectively. (D, E) Captured upconverted images of a stereo target at the distance $u$ and after the positional shift $\Delta u$, respectively. Notably, the scale bars in (B-E) correspond to the sizes of the recorded upconversion images at the camera. (F) Reconstructed stacked object. A physical view of the target is shown in the top-right corner. Here, $d_1$ denotes the object size, while $d_2$ and $d'_2$ denote the measured image sizes at the two respective positions.}
	\label{fig5}
\end{figure*}

In our scheme, the nonlinear optical pinhole converts weak MIR signals into the shorter-wavelength region, enabling detection with a silicon-based EMCCD that offers high spatial resolution and single-photon sensitivity. This configuration enables wide-field MIR tomographic imaging even under photon-starved conditions. From the recorded photon counts, the light flux into the upconverter is estimated to be 1.5 photons/pulse. The acquired volumetric dataset consists of 170 frames, with a depth step of 0.14 mm. An integration time of 10 s per frame is used to collect sufficient signal photons. Figure \ref{fig4}(B) shows the point cloud obtained by peak-searching along the temporal axis of the raw data. However, noise significantly degrades the ability to reconstruct object contours at this stage. To address this, a spatio-temporal correlation enhancement algorithm is applied to better extract and distinguish signal photons from noise \cite{Fang2023LSA}, producing a cleaner point cloud shown in Fig. \ref{fig4}(C). Additional edge-based denoising is then applied to generate the final 3D reconstruction in Fig. \ref{fig4}(D). As a result, a target with clear structure located at $u$ = 35 cm can be revealed across a wide field of view up to 6 cm, which is significantly larger than that in previous upconversion imagers \cite{Dam2012NP, Huang2022NC, Zeng2023LPR, Wang2023NC, Ge2023PRAppl, Mrejen2020LPR, Morales2021AP}. Notably, the object can be placed farther away to achieve an even wider field of view. Therefore,  high-contrast MIR imaging could have been realized with wide field of view and high detection sensitivity, despite of the intrinsically low light throughput of the pinhole.

\subsection{Depth-resolved MIR transmission imaging}
Finally, we proceed to investigating the photographic depth-resolving measurements, where the illumination is configured in the transmissive modality. The core principle is illustrated in Fig. \ref{fig5}(A). Specifically, the MIR pinhole imager, comprising the nonlinear converter and the camera, captures upconverted images with the object placed at distances $u$ and $u + \Delta u$. With the help of the object-image trigonometric relationship, the spatial dimensions and axial depth of the object can be recovered. As an example, the top-left corner of Fig. \ref{fig5}(A) shows a planar mask engraved with the acronym ``ECNU". The corresponding upconverted images at distances $u$ and $u + \Delta u$ are shown in Figs. \ref{fig5}(B) and (C), respectively. The object height $d_1$ remains constant and the image distance $v$ is fixed in the experiment. With a shift $\Delta u$ = 3 cm, the image heights $d_2$ and $d'_2$ differ at the two positions, and can be determined from the measured images. Using the formulas in Fig. \ref{fig5}(A), the vertical size of the letter ``E'' on the object plane is calculated to be about 5.98 mm, and its distance $u$ is estimated to be 13.97 cm. These measurements are consistent to the actual object size of 6 mm and the object distance of 14 cm.

Furthermore, we demonstrate the feasibility of photographic depth analysis for stacked objects. When visual parallax fails to distinguish relative size and depth, particularly when the object is positioned at a long distance, capturing two images at different object positions offers a fast and straightforward means of extracting structural information for stereoscopic reconstruction. Figure \ref{fig5}(D) shows an upconverted image of a stereo ``ECNU" mask at an initial object distance $u$, where the letters appear to differ in size. However, it is unclear whether this disparity arises from actual size differences or from object-to-image scaling. To resolve this ambiguity, the object is moved away from the detector by $\Delta u=3$ cm, and a second image is captured in Fig. \ref{fig5}(E). The reconstructed structure is presented in Fig. \ref{fig5}(F), alongside the ground-truth geometry shown in the top-right corner. The axial depth of the object spans nearly 6 cm, which underscores the unique advantage of pinhole imaging in accommodating extended depth without compromising image sharpness. Unlike conventional time-resolved imaging, this novel passive MIR depth measurement approach retrieves depth information without relying on the pulsed gating. The approach provides a simple, rapid, and data-efficient solution for reconstructing the depth information of the scene from just two images.

\section{Conclusion}
The presented work demonstrates a MIR lensless imaging system based on the nonlinear pinhole filtering. This method overcomes the inherent limitations of conventional lens-based systems, particularly in terms of DoF, field of view, and optical aberrations, offering an effective solution for high-sensitivity MIR imaging in complex environments. The core innovation lies in the use of a nonlinear optical pinhole, which enables image formation across an extended depth range while simultaneously supporting sensitive MIR signal detection. Notably, the shape and size of the optical aperture can be flexibly and precisely tailored, enhancing image quality and adaptability to diverse application requirements. Furthermore, the involved ultrashort optical gating allows the system to resolve depth information based on the arrival time of signal photons, enabling high-precision three-dimensional imaging. In addition, the intrinsic phase-matching filtering in the spatial and spectral domains favors suppressing ambient noises, which alleviates the complexity to make the camera obscura as required in conventional pinhole imaging.

In contrast to previous MIR upconversion imaging schemes \cite{Barh2019AOP, Dam2012NP, Huang2022NC, Zeng2023LPR, Wang2023NC}, the proposed approach introduces several distinct advantages. First, the system offers an effectively extended DoF, which allows clear image acquisition over a broad object distance range without being constrained to a specific focal plane. Second, the imager achieves distortion-free 3D tomographic imaging. This performance arises from the combination of the large acceptance angle of the CPLN-based upconverter and the extensive DoF provided by the nonlinear pinhole, enabling the spatial field of view to expand proportionally with object distance. Furthermore, the system is compatible with passive dual-view depth measurement, which allows the use of continuous-wave MIR illumination to offer greater flexibility in practical deployment.

In summary, we have devised and implemented for the first time to our knowledge a MIR nonlinear pinhole imaging architecture, which significantly enhances the DoF and field of view of the frequency upconversion imaging system. The nonlinear lensless design is not only fundamentally intriguing to complement the traditional linear pinhole imagers, but also opens new possibilities for optical imaging at extreme wavelengths. The use of a chirped-period nonlinear crystal in the upconversion stage enables broadband phase matching over a wide range of MIR wavelengths up to 5 $\mu$m without requiring active temperature control or adjustment of the pump configuration \cite{Huang2022NC, Fang2023LSA}. Furthermore, the presented paradigm is generic, and can be readily extended to other nonlinear media to facilitate lensless imaging in far-infrared or terahertz wavelengths \cite{Fandio2024OL,Rodrigo2021LPR}, where the fabrication of aberration-free lenses is challenging or the large-DoF operation is highly demanded in these spectral regions. Future explorations include the implementation of digital spatial mode modulation on the pump beam \cite{Wang2023NC}, allowing on-demand  dynamic tuning of pinhole properties to optimize the imaging performance. Furthermore, a multi-aperture encoding strategy can be used to implement computational coded imaging \cite{Horisaki2020OL}, which could improve the light throughput and reduce the acquisition time in low-light environments. With these advancements, the MIR nonlinear pinhole imaging paradigm could offer a disruptive tool for a wide range of applications including industrial process control, infrared machine vision, and night autopilot.

\vspace{24pt}
\noindent  {\fontfamily{phv}\selectfont 
\normalsize \textbf{Funding.} 
}
\noindent {\normalsize Shanghai Pilot Program for Basic Research (TQ20220104); National Natural Science Foundation of China (62175064, 62235019, 62035005); Innovation Program for Quantum Science and Technology (2023ZD0301000); Shanghai Municipal Science and Technology Major Project (2019SHZDZX01); Natural Science Foundation of Chongqing (CSTB2023NSCQ-JQX0011, CSTB2022TIAD-DEX0036);  China Postdoctoral Science Foundation (2024M760918, 2025T180224); Fundamental Research Funds for the Central Universities.}

\vspace{8pt}
\noindent  {\fontfamily{phv}\selectfont 
	\normalsize \textbf{Disclosures.} 
}
\noindent {\normalsize The authors declare no conflicts of interest.}

\vspace{8pt}
\noindent  {\fontfamily{phv}\selectfont 
	\normalsize \textbf{Data availability.} 
}
\noindent {\normalsize Data underlying the results presented in this paper are not publicly available at this time but may be obtained from the authors upon reasonable request.}

\vspace{8pt}
\noindent  {\fontfamily{phv}\selectfont 
	\normalsize \textbf{Supplemental document.} 
}
\noindent {\normalsize See Supplement 1 for supporting content.}


\begin{thebibliography}{100}

\bibitem{Ozcan2016ARBE} A. Ozcan, E. McLeod, ``Lensless imaging and sensing," \textit{Ann. Rev. Biomed. Eng.} \textbf{18}, 77-102 (2016).

\bibitem{Li2024FR} S. Li, Y. Gao, J. Wu, M. Wang, Z. Huang, S. Chen, L. Cao, ``Lensless camera: Unraveling the breakthroughs and prospects," \textit{Fundam. Res.} (2024).

\bibitem{Boominathan2022Optica} V. Boominathan, J. T. Robinson, L. Waller, A. Veeraraghavan, ``Recent advances in lensless imaging," \textit{Optica} \textbf{9}, 1-16 (2021).

\bibitem{Potter2024LPR} C. J. Potter, Z. Xiong, E. McLeod, ``Clinical and biomedical applications of lensless holographic microscopy," \textit{Laser Photon. Rev.} \textbf{18}, 2400197 (2024).

\bibitem{Antipa2018Optica} N. Antipa, G. Kuo, R. Heckel, B. Mildenhall, E. Bostan, R. Ng, L.Waller, ``DiffuserCam: lensless single-exposure 3D imaging," \textit{Optica} \textbf{5}, 1-9 (2018).

\bibitem{Hua2020IEEE TPAMI} Y. Hua, S. Nakamura, M. S. Asif, A. C. Sankaranarayanan, ``SweepCam-depth-aware lensless imaging using programmable masks," \textit{IEEE Trans. Pattern Anal. Mach. Intell} \textbf{42}, 1606–1617 (2020).

\bibitem{Sinha2017Optica} A. Sinha, J. Lee, S. Li, G. Barbastathis, ``Lensless computational imaging through deep learning," \textit{Optica} \textbf{4}, 1117-1125 (2017).

\bibitem{Wu2024LSA} X. Wu, N. Zhou, Y. Chen, J. Sun, L. Lu, Q. Chen, C. Zuo, ``Lens-free on-chip 3D microscopy based on wavelength-scanning Fourier ptychographic diffraction tomography," \textit{Light Sci. Appl.} \textbf{13}, 237 (2024).

\bibitem{Monakhova2020Optica} K. Monakhova, K. Yanny, N. Aggarwal, L. Waller, ``Spectral DiffuserCam: lensless snapshot hyperspectral imaging with a spectral filter array," \textit{Optica} \textbf{7}, 1298-1307 (2020).

\bibitem{Baek2022APLP} N. Baek, Y. Lee, T. Kim, J. Jung, S. A. Lee, ``Lensless polarization camera for single-shot full-stokes imaging," \textit{APL Photon.} \textbf{7}, 116107 (2022).

\bibitem{Liang2020RPP} J. Liang, ``Punching holes in light: recent progress in single-shot coded-aperture optical imaging," \textit{Rep. Prog. Phys.} \textbf{83}, 116101 (2020).

\bibitem{Cieslak2016RM} M. J. Cie\'{s}lak, K. A. A. Gamage, R. Glover, ``Coded-aperture imaging systems: past, present and future development-A review," \textit{Radiat. Meas.} \textbf{92}, 59-71(2016).

\bibitem{Boominathan2016IEEE SPM} V. Boominathan, J. K. Adams, M. S. Asif, B. W. Avants, J. T. Robinson, R. G. Baraniuk, A. C. Sankaranarayanan, A. Veeraraghavan, ``Lensless Imaging: A computational renaissance," \textit{IEEE Signal Process. Mag.} \textbf{33}, 23-35 (2016).

\bibitem{Gong2009APL}  W. Gong, P. Zhang, X. Shen, S. Han, ``Ghost `pinhole' imaging in Fraunhofer region," \textit{Appl. Phys. Lett.} \textbf{95}, 071110 (2009).

\bibitem{Wang2015OC} Z. Wang, W. Guo, ``Coded pinhole lens imaging," \textit{OSA Contin.} \textbf{1}, 64 (2018).

\bibitem{Franke1979AO} J. M. Franke, ``Field-widened pinhole camera," \textit{Appl. Opt.} \textbf{18}, 2913-2914 (1979).

\bibitem{Lindberg1970AHES} D. C. Lindberg, ``The theory of pinhole images in the fourteenth century," \textit{Arch. Hist. Exact Sci.} \textbf{6}, 299-325 (1970).

\bibitem{Young1989PT} M. Young, ``The pinhole camera: imaging without lenses or mirrors," \textit{Phys. Teach.} \textbf{27}, 648-655 (1989).

\bibitem{Straker1981AHES} S. Straker, ``Kepler, Tycho, and the `Optical Part of Astronomy': the genesis of Kepler's theory of pinhole image," \textit{Arch. Hist. Exact Sci.} \textbf{24}, 267-293 (1981).

\bibitem{Biri2011IEEE TPS} S. Biri, E. Takacs, R. Racz, L. T. Hudson, J. Palinkas, ``Pinhole X-Ray camera photographs of an ECR ion source plasma," \textit{IEEE Trans. Plasma Sci.} \textbf{39}, 2494-2495 (2011).

\bibitem{Gallas1965JSMPTE} A. H. Gallas, C. A. Gilbert, A. B. Hitterdal, ``Pinhole optics and simulators," \textit{J. Soc. Motion Pict. Telev. Eng.} \textbf{74}, 321-323 (1965).

\bibitem{Newman1966AO} P. A. Newman, V. E. Rible, ``Pinhole array camera for integrated circuits," \textit{Appl. Opt.} \textbf{5}, 1225-1228 (1966).

\bibitem{Young1971AO} M. Young, ``Pinhole optics," \textit{Appl. Opt.} \textbf{10}, 2763-2767 (1971).

\bibitem{Vodopyanov2020Book} K. L. Vodopyanov, \textit{Laser-based Mid-infrared Sources and Applications}, Wiley (2020).

\bibitem{Rogalski2005RPP} A. Rogalski, ``HgCdTe infrared detector material: history, status and outlook," \textit{Rep. Prog. Phys.} \textbf{68}, 2267 (2005).

\bibitem{Wang2019Small} P. Wang, H. Xia, Q. Li, F. Wang, L. Zhang, T. Li, P. Martyniuk, A. Rogalski, W. Hu, ``Sensing infrared photons at room temperature: from bulk materials to atomic layers," \textit{Small} \textbf{15}, 1904396 (2019).

\bibitem{Taylor2023Optica} G. G. Taylor, A. B. Walter, B. Korzh, B. Bumble, S. R. Patel, J. P.Allmaras, A. D. Beyer, R. O'Brient, M. D. Shaw, E. E. Wollman, ``Low-noise single-photon counting superconducting nanowire detectors at infrared wavelengths up to 29 $\mu$m," \textit{Optica} \textbf{10}, 1672-1678 (2023).

\bibitem{Keuleyan2011NP} S. Keuleyan, E. Lhuillier, V. Brajuskovic, P. G. Sionnest, ``Mid-infrared HgTe colloidal quantum dot photodetectors," \textit{Nat. Photon.} \textbf{5}, 489–493 (2011).

\bibitem{Bullock2018NP} J. Bullock, M. Amani, J. Cho, Y. Z. Chen, G. H. Ahn, V. Adinolfi, V. R. Shrestha, Y. Gao, K. B. Crozier, Y. L. Chueh, A. Javey, ``Polarization-resolved black phosphorus/molybdenum disulfide mid-wave infrared photodiodes with high detectivity at room temperature," \textit{Nat. Photon.} \textbf{12}, 601-607 (2018).

\bibitem{Guo2018NM} Q. Guo, R. Yu, C. Li, S. Yuan, B. Deng, F. J. Garc\'{i}a de Abajo, F. Xia, ``Efficient electrical detection of mid-infrared graphene plasmons at room temperature," \textit{Nat. Mater.} \textbf{17}, 986-992 (2018).

\bibitem{Peng2021SA} M. Peng, R. Xie, Z. Wang, P. Wang, F. Wang, H. Ge, Y. Wang, F. Zhong, P. Wu, J. Ye, Q. Li, L. Zhang, X. Ge, Y. Ye, Y. Lei, W. Jiang, Z. Hu, F. Wu, X. Zhou, J. Miao, J. Wang, H. Yan, C. Shan, J. Dai, C. Chen, X. Chen, W. Lu, W. Hu, ``Blackbody-sensitive room-temperature infrared photodetectors based on low-dimensional tellurium grown by chemical vapor deposition," \textit{Sci. Adv.} \textbf{7}, eabf7358 (2021).

\bibitem{Liu2021LSA} C. Liu, J. Guo, L. Yu, J. Li, M. Zhang, H. Li, Y. Shi, D. Dai, ``Silicon/2D-material photodetectors: from near-infrared to mid-infrared," \textit{Light Sci. Appl.} \textbf{10}, 123 (2021).

\bibitem{Barh2019AOP} A. Barh, P. J. Rodrigo, L. Meng, C. Pedersen, P. Tidemand-Lichtenberg, ``Parametric upconversion imaging and its applications," \textit{Adv. Opt. Photon.} \textbf{11}, 952 (2019).

\bibitem{Dam2012NP} J. S. Dam, P. Tidemand-Lichtenberg, C. Pedersen, ``Room-temperature mid-infrared single-photon spectral imaging," \textit{Nat. Photon.} \textbf{6}, 788 (2012).

\bibitem{Huang2022NC} K. Huang, J. Fang, M. Yan, E Wu, H. Zeng, ``Wide-field mid-infrared single-photon upconversion imaging," \textit{Nat. Commun.} \textbf{13}, 1077 (2022).

\bibitem{Zeng2023LPR} X. Zeng, C. Wang, H. Wang, Q. Lin, Z. Chen, X. Lu, M. Zheng, J. Liang, Y. Cai, S. Xu, J. Li, ``Tunable Mid-Infrared Detail-Enhanced Imaging With Micron-Level Spatial Resolution and Photon-Number Resolving Sensitivity," \textit{Laser Photon. Rev.}  \textbf{17}, 2200446 (2023).

\bibitem{Wang2023NC} Y. Wang, K. Huang, J. Fang, M. Yan, E Wu, H. Zeng, ``Mid-infrared single-pixel imaging at the single-photon level," \textit{Nat. Commun.} \textbf{14}, 1073 (2023).

\bibitem{Ge2023PRAppl} Z. Ge, Z. Han, Y. Liu, X. Wang, Z. Zhou, F. Yang, Y. Li, Y. Li, L. Chen, W. Li, S. Niu, B. Shi, ``Midinfrared up-conversion imaging under different illumination conditions," \textit{Phys. Rev. Appl.} \textbf{20}, 054060 (2023).

\bibitem{Mrejen2020LPR} M. Mrejen, Y. Erlich, A. Levanon, H. Suchowski, ``Multicolor Time-Resolved Upconversion Imaging by Adiabatic Sum Frequency Conversion," \textit{Laser Photon. Rev.} \textbf{14}, 2000040 (2020).

\bibitem{Morales2021AP} R. C. Morales, D. Rocco, L. Xu, V. F. Gili, N. Dimitrov, L. Stoyanov, Z. Ma, A. Komar, M. Lysevych, F. Karouta, A. Dreischuh, H. H. Tan, G. Leo, C. D. Angelis, C. Jagadish, A. E. Miroshnichenko, M. Rahmani, D. N. Neshev, ``Infrared upconversion imaging in nonlinear metasurfaces," \textit{Adv. Photon.} \textbf{3}, 036002-036002 (2021).

\bibitem{Zheng2013NP} Q. Zheng, H. Zhu, S. C. Chen, C. Tang, E. Ma, X. Chen, ``Frequency-upconverted stimulated emission by simultaneous five-photon absorption," \textit{Nat. Photon.} \textbf{7}, 234-239 (2013).

\bibitem{Knez2020LSA} D. Knez, A. M. Hanninen, R. C. Prince, E. O. Potma, D. A. Fishman, ``Infrared chemical imaging through non-degenerate two-photon absorption in silicon-based cameras," \textit{Light Sci. Appl.} \textbf{9}, 125 (2020).

\bibitem{Tomasi2015NCS} C. Tomasi, ``A simple camera model," \textit{Notes Comput. Sci.} \textbf{527}, (2015).

\bibitem{Liu2020AO} L. Liu, L. Shi, A. Cao, H. Pang, W. Yan, Y. Pang, L. Xue, W. Liu, Q. Deng, ``Design and experiment of a soft-edge aperture with high light energy utilization efficiency and uniformity," \textit{Appl. Opt.} \textbf{59}, 5348-5357 (2020).

\bibitem{Gariepy2015NC} G. Gariepy, N. Krstaji\'{c}, R. Henderson, C. Li, R. R. Thomson, G. S. Buller, B. Heshmat, R. Raskar, J. Leach, D. Faccio, ``Single-photon sensitive light-in-flight imaging," \textit{Nat. Commun.} \textbf{6}, 6021 (2015).

\bibitem{Fang2023LSA} J. Fang, K. Huang, E Wu, M. Yan, H. Zeng, ``Mid-infrared single-photon 3D imaging," \textit{Light Sci. Appl.} \textbf{12}, 144 (2023).

\bibitem{Rodrigo2021LPR} P. J. Rodrigo, L. H$\o$gstedt, S. M. M. Friis, L. R. Lindvold, P. Tidemand-Lichtenberg, C. Pedersen, ``Room-temperature, high-SNR upconversion spectrometer in the 6–12 $\mu$m region," \textit{Laser Photon. Rev.} \textbf{15}, 2000443 (2021).

\bibitem{Fandio2024OL} D. J. J. Fandio, A. Vishnuradhan, E. K. Yalavarthi, W. Cui, N. Couture, A. Gamouras, J. M. M\'{e}nard, ``Zeptojoule detection of terahertz pulses by parametric frequency upconversion," \textit{Opt. Lett.} \textbf{49}, 1556 (2024).

\bibitem{Horisaki2020OL} R. Horisaki, Y. Okamoto, J. Tanida, ``Deeply coded aperture for lensless imaging," \textit{Opt. Lett.} \textbf{45}, 3131 (2020).
  
\end{thebibliography}
\end{document}